\documentstyle[12pt]{article}
\begin{document}

\title{Scanning of hadron cross--section at DA$\Phi$NE by analysis of
the initial--state radiative events}
\author{M.Konchatnij, N.P.Merenkov}
\date{}
\maketitle
\begin{center}

\small{\it{National Sience Centre  \\
"Kharkov Institute of Physics and Technology"}}
\end{center}
\vspace{0.5cm}
\begin{abstract}
The initial--state radiative events in electron--positron annihilation
into hadrons at DA$\Phi$NE have been considered. The corresponding
cross--section with the full first order radiative corrections has been
calculated. The analytical calculations take into account the realistic
angular acceptance and energy cut of DA$\Phi$NE photon detector.
\end{abstract}

It is the general viewpoint at present that the dominant error in SM fits
arises due to fairly large uncertainity in hadron contribution to vacuum
polarization (HVP). The limited knowledge of the HVP affects also the
precise determination of the muon anomalous magnetic moment [1,2]. The
main problem is connected with the low and intermediate energies because
in these regions HVP cannot be computed by means QCD. The only possibility
is to reconstruct it by dispersion intrgral using the
measured total cross--section of the process $e^+e^-\rightarrow hadrons $
at continuously varying energies. Therefore, there exists an eminent
physical reason to scan the total hadronic cross--section in the region up
to a few GeV.

For this goal one can exploit the high luminosity of oncoming $e^+e^-$
colliders which operate at fixed energy and use the process of radiative
return to lower energies due to initial--state photon emission [3--5]:
\begin{equation}\label{1}
e^-(p_1) + e^+(p_2) - \gamma(k) \rightarrow \gamma^*(q)\rightarrow H(q) \
, \ \ q=p_1+p_2-k \ ,
\end{equation}
where $H$ denotes all final hadrons, $\gamma^*$ is the intermediate heavy
photon. We write the final photon on the left side of (1) to emphasize
that it radiates from the initial state. The total hadronic cross--section
$\sigma_t$ of the process (1), that defines the imaginary part of the HVP,
depends on the heavy photon virtuality $$ q^2 = s(1-x) \ , \ \  s =
4\varepsilon^2 \ , \ \ x = \frac{\omega}{\varepsilon} \ , $$ where $\omega
(\varepsilon)$ is the photon (electron) energy.  Therefore,
by measuring the photon energy fraction $x$ one can
extract the distribution $\sigma(s(1-x)).$

In the case of final--state radiation process
\begin{equation}\label{2}
e^-(p_1) + e^+(p_2) \rightarrow\gamma^*(p_1+p_2)\rightarrow\gamma(k) + H(q)
\end{equation}
the quantity $\sigma_t$ depends on $s,$ while the hadron invariant mass
$q^2$ leaves the same. Cosequently, in order to scan the total hadronic
cross--section it needs to use only events with initial--state radiation.
The events with final--state radiation have to be considered as a
background. In Ref.[5] authors suggested to restrict the angular hadron
phase space to get rid of the final--state radiation background. On our
opinion the such procedure is not adequate to main goal, because the
exctracted in this case experimental cross--section must depends on the
restriction parameters, whereas the quantity $\sigma_t$ (which enters into
dispersion integral) depends, by definition, on the heavy photon invariant
mass only.

In present work we calculate the quantity $\sigma_t$ for the process (1)
and radiative correction to it, using the realistic conditions of the
tagged photon detector (PD) at DA$\Phi$NE collider. The angular acceptance
of the DA$\Phi$NE PD covers all phase space except for a two symmetrical
cones with the opening angle $2\theta_0$ along both, the electron and the
positron beam directions. Such kind of angular acceptance is just opposite
to one used in [4] where the PD has covered narrow cone along the electron
(or the positron) beam direction. Besides, the realistic DA$\Phi$NE PD
selects the events with only one hard photon hitting it, and the
coresponding energy cutoff parameter is $\Delta .$ Under the DA$\Phi$NE
conditions
\begin{equation}\label{3}
\theta_0 = 10^{\circ}   \ , \ \ \ \Delta = 4\cdot 10^{-2} \ ,
\end{equation}
and the radiative corrected total hadronic cross--section will depend on
both, the angular and the energy cutoff parameters.

The Born cross--section of the process (1) can be written as follows
[4,6]
\begin{equation}\label{4}
d\sigma^B =
\int\limits_{\Omega(\theta_0)}^{}\sigma(q^2)\frac{\alpha}{2\pi^2}\frac
{(s+t_1)^2+(s+t_2)^2}{t_1t_2}\frac{d^3k}{s\omega} \ ,
\end{equation}
where $\Omega(\theta_0)$ covers the angular acceptance of PD, and $t_{1,2}
=-2(kp_{1,2}).$ The normalizaion cross--section under integral sign on the
right side of Eq.(4) can be expressed in terms of the ratio $R$ of the
total hadronic and muonic cross--sections:
\begin{equation}\label{5}
\sigma(q^2) =
\frac{4\pi\alpha^2}{3q^2|1-\Pi(q^2)|^2}R(q^2)\bigl(1+\frac{2\mu^2}{q^2}
\bigr)\sqrt{1-\frac{4\mu^2}{q^2}} \ , \ \
 R(q^2) = \frac{\sigma_t(e^+e^-\rightarrow hadrons)}{\sigma_t(e^+e^-
\rightarrow \mu^+\mu^-)} \ ,
\end{equation}
where $\mu $ is the muon mass. The lepton contribution into vacuum
polarization $\Pi(q^2)$ is the known function [1] and will not be specific
here.
The angular integration of Eq.(4) gives \begin{equation}\label{6}
d\sigma^B = \frac{\alpha}{2\pi}\sigma(q^2)2\bigl[\frac{1+(1-x)^2}{x}\ln
\frac{1+\cos\theta_0}{1-\cos\theta_0} - x\cos\theta_0\bigr]dx \ .
\end{equation}
In the Born approximation the cross--section of the process (1) depends on
the angular cutoff parameter only. Looking at Eq.(6) we see that the
measurement of that cross--section at different values of the tagged
photon energy fraction $x$ allows to extract the quantity $\sigma(e^+e^-
\rightarrow hadrons)$ at different effective collision energies $s(1-x).$

The high precision measurement of the total hadronic cross--section
requires to adequate theoretical calclations. These last have to take into
account at least the first order radiative correction (RC). The RC to
$d\sigma^B$ includes the contributions due to aditional virtual and real
soft ( with the energy less then $\varepsilon\Delta \ , \omega <
\varepsilon\Delta$) photon emission in all angular phase space as well as
due to hard $(\omega > \varepsilon\Delta)$ photon emission in the region
where the PD does not record it.

When calculating the RC to the Born cross--section we suggest that
$\sigma(q^2)$ is the
flat enough function of its argument, such that the conditions
\begin{equation}\label{7}
\frac{\Delta}{\sigma(q^2)}\frac{d\sigma}{d\ln(q^2)} \ll 1 \ ,  \ \
\frac{\theta_0^2}{\sigma(q^2)}\frac{d\sigma}{d\ln(q^2)} \ll 1
\end{equation}
are satisfied. These conditions permit to apply the soft photon
approximatiom and the quasireal electron method [8] for a description of
only wide resonance contributions into cross--section.

The virtual and soft photon corretions can be computed using the results
of work [7] where one--loop corrected Compton tensor with a heavy photon
was calculated for the scattering channel. In order to reconstruct the
corresponding results for the annihilation channel it is enough to change
(in accordance with the notation used here)
\begin{equation}\label{8}
p_2\rightarrow-p_2\ , \ \ u\rightarrow s \ , \ \ t\rightarrow t_1
\end{equation}
in all formulae of the Ref.[7].

Thus, the contribution of virtual and soft photon emission into the RC to
Born cross--section can be written as follows
\begin{equation}\label{9}
d\sigma^{V+S} =
\int\limits_{\Omega(\theta_0)}^{}\sigma(q^2)\frac{\alpha^2}{4\pi^3}
\bigl[\rho\frac
{(s+t_1)^2+(s+t_2)^2}{t_1t_2} + T \bigl]\frac{d^3k}{s\omega} \ ,
\end{equation}
\begin{equation}\label{10}
\rho = 4(l_s-1)\ln\Delta + 3(l_s + \ln(1-x)) - \frac{\pi^2}{3} -
\frac{9}{2} \ , \ \ l_s = \ln\frac{s}{m^2} \ ,
\end{equation}
\begin{equation}\label{11}
T = \frac{3}{2}T_g - \frac{1}{8q^2}\{T_{11}(s+t_1)^2 + T_{22}(s+t_2)^2
+(T_{12}+T_{21})[s(s+t_1+t_2)-t_1t_2]\} \ ,
\end{equation}
where $m$ is the electron mass. For quantities $T_g$ and $T_{ik}$ see [7],
bearing in mind substitution (8). It needs to note only that under the
DA$\Phi$NE conditions $|t_{1,2}|_{min}\approx\varepsilon^2\theta_0^2
\gg m^2,$ therefore one have to omitt terms proportional to $m^2$ in both
$T_g$ and $T_{ik}.$

The Born--like structure, that contains multiplier $\rho,$ on the right
side of Eq.(9) absorbs all infrared singularities via quantity
$\ln\Delta.$ In the limiting case
$$|t_1|=2\varepsilon^2(1-c)\approx\varepsilon^2\theta_0^2\ll s, \
|t_2|\ ; \ \ t_2=-sx\ , \ \ q^2 = s(1-x) \ , $$
which corresponds to events with the tagged photon detected very close to
the cutoff angle $\theta_0$ along the electron beam direction, the
expression into parenthesis on the right side of Eq.(9) reads
\begin{equation}\label{12}
2\rho\frac{1+(1-x)^2}{x^2(1-c)} +
\frac{2}{x(1-c)}\bigl\{\frac{1+(1-x)^2}{x}\bigl[\ln(1-x)\ln\frac{x^2(1-c)}
{2(1-x)}-2f(x)\bigr] + \frac{2-x^2}{2x}\bigr\}\ ,
\end{equation}
$$f(x) = \int\limits_0^x\frac{dz}{z}\ln(1-z) \ , \ \ c = \cos\theta \ , $$
where $\theta$ is the angle between vectors $\vec k$ and $\vec p_1.$
For events with recorded photon very close to the cutoff angle
$\theta_0$ along the positron beam direction it needs to change $c$ in (12)
by $-c.$

To compute the RC due to invisible hard photon radiated along the
electron beam direction inside the cone with the opening angle $2\theta_0$
we can use the quasireal electron method [8]. In accordance with this
method the corresponding contribution into cross--section has a form
\begin{equation}\label{13}
d\sigma^H_1 = \int d\sigma^B(p_1-k_1,k,p_2)dW_{p_1}(k_1)
\end{equation}
where $d\sigma^H_1$ is the cross--section of the process
\begin{equation}\label{14}
e^-(p_1)+e^+(p_2)-\gamma(k)-\gamma(k_1)\rightarrow\gamma^*
\rightarrow H(q)
\end{equation}
provided the additional hard photon $\gamma(k_1)$ is emitted along the
electron beam direction.

The expression for the radiation probability $dW(k_1)$ is well known [8],
and under the DA$\Phi$NE conditions it may be written as folows
\begin{equation}\label{15}
dW(k_1) =
\frac{\alpha}{2\pi}P\bigl(1-z,\ln\frac{\varepsilon^2\theta_0^2}{m^2}
\bigr)dz \ , \ \ P(1-z,L) = \frac{1+(1-z)^2}{z}L - \frac{2(1-z)}{z} \ ,
\end{equation}
where $z$ is the energy fraction of invisible photon, and we use
approximation: $2(1-\cos\theta_0) = \theta_0^2$ in argument of logarithm.

The shifted Born cross--section on the right side of Eq.(13) is defined by
the formula
\begin{equation}\label{16}
d\sigma^B(p_1-k_1,k,p_2) = \int\limits_{\Omega(\theta_0)}^{}\frac{\alpha}
{2\pi}\sigma(q_1^2)\frac{(1-z)^2(s+t_1)^2+((1-z)s+t_2)^2}{(1-z)^2t_1t_2}\frac
{d^3k}{s\omega}\ ,
\end{equation}
$$   q_1 = (1-z)p_1+p_2-k \ .$$
After integration over the invisible photon energy fraction $z$ on the
right side of Eq.(13) we derive
\begin{equation}\label{17}
d\sigma^H_1 = \frac{\alpha}{2\pi}\int\limits_{\Delta}^{z_m}dzP\bigl(1-z,
\ln\frac{\varepsilon^2\theta_0^2}{m^2}\bigr)d\sigma^B(p_1-k_1,k,p_2) \ ,
\end{equation}
where the upper limit of integration is defined by condition $q_1^2 \geq
4m_{\pi}^2\ (m_{\pi}$ is the pion mass) and reads
$$z_m = \frac{2(1-x-\delta)}{2-x(1-c)} \ , \ \ \delta = \frac{4m_{\pi}^2}
{s} \ . $$

The corresponding expression for $d\sigma^H_2$, when the additional
invisible hard photon is emitted along the positron beam direction, can be
obtained from Eq.(17) by substitution $p_1\leftrightarrow p_2$ in
$d\sigma^B$ and $c\rightarrow -c$ in $z_m.$ Because the angular acceptance
of the DA$\Phi$NE PD is symmetrical respect to change $c\rightarrow -c,\ \
d\sigma^H_1 = d\sigma^H_2,$ and the full RC to the Born cross--section
reads
\begin{equation}\label{18}
d\sigma^{RC} = d\sigma^{V+S} + 2d\sigma^H_1 \ .
\end{equation}

It is useful to rewrite the function $P(1-z,L)$ where $L =
l_s+\ln\frac{\theta_0^2}{4},$ that enters into $d\sigma^H_1$ in the
following form
\begin{equation}\label{19}
P(1-z,L) = P_1(1-z,L) - 2G -
\delta(z)\bigl[\bigl(\frac{3}{2}+2\ln\Delta\bigr)L - 2\ln\Delta\bigr] \ ,
\end{equation}
$$P_1(1-z,L) = \bigl[\frac{1+(1-z)^2}{z}\theta(z-\Delta)+\delta(z)\bigl(
\frac{3}{2}+2\ln\Delta\bigr)\bigr]L \ , \ G =
\frac{1-z}{z}\theta(z-\Delta)+\delta(z)\ln\Delta \ , $$
where the quantity $(\alpha/2\pi)P_1(y,L)$ is the well known first
order electron structure function [9], and simultaniously suppose the
lower limit of $z$--integration in $d\sigma^H_1$ to be equal to zero. Then
the measured cross--section of the process (1) under the DA$\Phi$NE
conditions can be written as follows \begin{equation}\label{20} d\sigma =
d\sigma^B(p_1,k,p_2)\bigl\{1+\frac{\alpha}{2\pi}\bigl[(3+4\ln\Delta)\ln\frac
{4}{\theta_0^2}+3\ln(1-x) - \frac{\pi^2}{3}-\frac{9}{2}\bigr]\bigr\}
\end{equation}
$$+\frac{\alpha}{2\pi}\bigl\{\frac{\alpha}{2\pi^2}\int\limits_{\Omega
(\theta_0)}^{}\sigma(q^2)T\frac{d^3k}{s\omega} + 2\int\limits_0^{z_m}dz
[P_1(1-z,L)-G]d\sigma^B((1-z)p_1,k,p_2)\bigr\} \ .
$$
The term containing the product of logarithms of the energy and the
angle cutoff parameter arises because we do not permit for the
additional hard photon to appear inside PD. We can resum the main
contributions on the right side of Eq.(20) in all orders of perturbation
theory and write the master formula in the form
$$d\sigma = \int dz_1\int
dz_2d\sigma^B(z_1,z_2)\{D(z_1,L)D(z_2,L)-\frac{\alpha}{2\pi}[\delta(1-z_1)
G(1-z_2)+\delta(1-z_2)G(1-z_1)]\bigr\}\Theta_{12}$$
\begin{equation}\label{21}
+d\sigma^B(1,1)\bigl[\frac{\alpha}{2\pi}\bigl(3\ln(1-x)-\frac{\pi^2}{3}-\frac
{9}{2}\bigr) + \exp{(\beta
l_s)}(1-\exp{\bigl(\beta\ln\frac{\theta_0^2}{4}\bigr)})\bigr]
\end{equation}
$$+\frac{\alpha^2}{4\pi^2}\int\limits_{\Omega(\theta_0)}\sigma(q^2)T\frac
{d^3k}
{s\omega} \ , \ \ d\sigma^B(z_1,z_2) = d\sigma^B(z_1p_1,k,z_2p_2) \ , \ \
\beta = \frac{2\alpha}{\pi}\bigl(\frac{3}{2}+\ln\Delta\bigr) \ , $$
where D(z,L) is the full electron sructure function.
Teta--function $\Theta_{12}$ under integral sign defines the
integration limits over variables $z_1$ and $z_2$ provided
$$(z_1p_1+z_2p_2+k)^2 \geq 4m_{\pi}^2 \ .$$
The corresponding limits of integration can be written as follows
\begin{equation}\label{22}
1 > z_2 > \frac{2\delta+z_1x(1-c)}{2z_1-x(1+c)} \ , \ \
1 > z_1 > \frac{2\delta+x(1+c)}{2-x(1-c)} \ .
\end{equation}
The cross--section (21) corresponds to such event selection when one hard
photon with the energy fraction $x$ hitting PD and accompaning with the
arbitrary number of soft photons with the energy fraction up to $\Delta$
for every ones inside PD is included as an event. If we want select events
when the energy fraction of all soft photons inside PD does not exeed
$\Delta$ we have to change $\beta$ by
$\bar\beta = \frac{\alpha}{2\pi}\bigl(\frac{3}{4} -C+\ln\Delta\bigr),$
where $C$ is the Euler costant and write
$$\frac{1}{\Gamma\bigl(1+\frac{\alpha}{2\pi} l_s\bigr)} -
\frac{\exp\bigl(\bar\beta
\ln\frac{\theta_0^2}{4}\bigr)}{\Gamma\bigl(1+\frac{\alpha}{2\pi}L\bigr)}$$
instead of $1-\exp\bigl(\beta\ln\frac{\theta_0^2}{4}\bigr)$ on the second
line of Eq.(21).

The master formula (21) takes into account only photonic RC. It can be
genegalized in such a way to include also the leading corrections due to
electron--positron pair production. We will assume that inside PD only
soft pairs (with the energy fraction less than $\Delta$) can be present,
while outside PD both, the soft and hard ones. In this case the
corresponding generalization can be carried out by insertion of effective
electromagnetic coupling [9] instead of $\alpha l_s$ and $\alpha L$
\begin{equation}\label{23}
\alpha l_s\rightarrow -3\pi\ln\bigl(1-\frac{\alpha}{3\pi}l_s\bigr) \ , \ \
\alpha L\rightarrow -3\pi\ln\bigl(1-\frac{\alpha}{3\pi}L\bigr) \ ,
\end{equation}
in the electron structure functions and exponents on the right side of
Eq.(21). Besides this we have to represent the electron structure function
as a sum of nonsinglet and singlet parts [9]. Note that the nonsinglet part
can be written in iterative as well as in exponentiated form, whereas the
singlet one has only iterative form. For the electron structure functions
see [9,10].

To extract the radiative corrected hadronic cross--section from
the corresponding experimental data of DA$\Phi$NE collider with a few
tenth percent accuracy it is enough to use Eq.(20). In this case we can
expand the quantity $\sigma(q^2)$ as
\begin{equation}
\sigma(q^2) = \sigma_0(q^2) + \frac{\alpha}{2\pi}\sigma_1(q^2)
\end{equation}
and obtain $\sigma_0$ and $\sigma_1$ by application of the simple iterative
procedure to Eq.(20) bearing in mind that the cross--section on the left
side of this equation is measured by experiment. The corresponding
equation in zero approximation
\begin{equation}\label{25}
d\sigma^{exp} = \frac{\alpha}{2\pi^2}\int\limits_{\Omega(\theta_0)}^{}
\sigma_0(q^2)\frac{(s+t_1)^2+(s+t_2)^2}{t_1t_2}\frac{d^3k}{s\omega}
\end{equation}
allows to extract the dependence $\sigma_0(q^2)$ in wide interval of
variable $q^2.$ In the first approximation the equation for $\sigma_1$
reads
$$
\int\limits_{\Omega(\theta_0)}^{}\biggl\{\biggl[\sigma_1(q^2) +
\sigma_0(q^2)\bigl[(3+4\ln\Delta)\frac{4}{\theta_0^2}+ 3\ln(1-x)
-\frac{\pi^2}{3}-\frac{9}{2}\bigr]\biggr]\frac{(s+t_1)^2+(s+t_2)^2}{t_1t_2}
$$
\begin{equation}\label{26}
+\sigma_0(q^2)T + 2\int\limits_0^{z_m}\sigma_0(q_1^2)[P_1(1-z,L)-G]
\frac{(1-z)^2(s+t_1)^2+((1-z)s+t_2)^2}{(1-z)^2t_1t_2}dz\biggr\} = 0 \ .
\end{equation}
This equation can be solved numerically respect to function
$\sigma_1(q^2).$

Authors thank A.B. Arbuzov, V. Khoze, G. Pancheri and L.
Trentadue for fruitful discussion and critical remarks. This work
supported in part (N.P.M) by INTAS Grant 93--1867 ext. N.P.M. thanks INFN
and the National Laboratory in Frascati for the hospitality. \\

\hspace{0.4cm}
{\large{References}} \\
\hspace{0.4cm}
\begin{enumerate}
\item D.H. Brown, W.A. Worstell, Phys. Rev. {\bf D 54} (1996) 3237;
S. Eidelman et al., Z. Phys. {\bf C 67} (1995) 585.
\item F. Jegerlehner. Transparencies of the Workshop on "Hadron production
cross--sections at DA$\Phi$NE", v.1, p.81. Karlsrue, Germany, November
1 and 2, 1996. Editors: U.V. Hagel, W. Kluge, J. K$\ddot u$hn, G.
Pancheri.
\item M.W. Krasny, W. Placzek, H. Spiesberger, Z. Phys. {\bf C 53} (1992)
687; D. Bardin, L. Kalinovskaya, T. Riemann, Z. Phys. {\bf C 76} (1997)
425; H. Anlauf, A.B. Arbuzov, E.A. Kuraev and N.P. Merenkov, Phys. Rev.
{\bf D 59} (1999) 014003.
\item A.B. Arbuzov, E.A. Kuraev, N.P. Merenkov, L. Trentadue, JHEP 12
(1998) 009.
\item S. Binner, J.H. K$\ddot u$hn and K. Melnikov,
HEP--PH/9902399.
\item V.N. Baier and V.A. Khoze, Sov. Phys. JETP, {\bf
21} (1965) 1145.
\item E.A. Kuraev, N.P. Merenkov, V.S. Fadin, Sov. J.
Nucl. Phys.  {\bf 45} (1987) 486.
\item V.N. Baier, V.S. Fadin, V.A. Khoze, {\bf B 65} (1973) 381.
\item E.A. Kuraev, V.S. Fadin, Sov. J. Nucl. Phys. {\bf 41} (1985) 466.
\item M. Skrzypek, Acta Phys. Pol. {\bf B 23} (1992) 135; S. Jadach,
M. Skrzypek and B.F.L. Ward, Phys. Rev. {\bf D 47} (1993) 3733.

\end{enumerate}

\end{document}